\documentstyle[12pt,epsfig]{article}                        
\newcommand{\ber}{\begin{eqnarray}}
\newcommand{\eer}{\end{eqnarray}}
\newcommand{\bea}{\begin{equation}}
\newcommand{\eea}{\end{equation}}
\newcommand{\del}{\partial}
\begin{document}
\title{\bf  Feynman-Kac path integral approach for the energy 
spectrum of many boson systems }      
\author {\bf S. Datta  \\
 S. N. Bose National Centre for Basic Sciences, India} 
\maketitle
\begin{abstract}
We study the ground and excited states of weakly interacing Bose gases ( with 
positive and negative scattering lengths) in connection with Bose Einstein 
condensation to test the validity of using the mean field theory and Born 
approximation. They behave as new quantum fluids ( a gas in the weak limit and 
a liquid in the dense limit and we study their many body physics in the dilute 
limit within the realistic potential model ( Morse type) by Feynman-Kac
path integral technique . Within numerical limitations, this method is exact in priciple and turns out to be a better alternative to GP as all the ground and 
excited states properties can be calculated in a much simpler way.
\end{abstract}
\newpage
\section{Introduction}
With the experimental realization of Bose Einstein Condensation in alkali 
gases[1], the study of many boson systems has become an area of active research 
interst. Previous numerical procedures are based on mean field theory like 
Gross Pitaevski [2] etc. They seem to work well for ground state properties but
turn out to be approximate as they fail to include correlations in the many 
body theory. Investigations of effects beyond mean field theory is an
important task and makes the many boson systems interesting even from the many 
body perspective[3]. Moreover  earlier calculations with $\delta$ function 
potential do not solve the many body problem exacxtly but only within a 
perturbation theory as in a system of Bose gas with $\delta $ function 
potential, particles do not collide [4]. As a result, speculations 
( particularly excitation frequencies  etc ) based on these methods differ 
drastically for different experimental modes. So an alternative to GP was 
necessary  which can describe the effect of interaction in a more reliable way 
and predict the excitation frequencies  and other properties more accurately.
Eventhough  Monte Carlo techniques are slow, computationally expensive
and faces sign problem for fermionic systems, these are the only numerical
techniques available for these kind of many body systems, which are exact
and can include corrlations in a reliable way. We also test the validity
of Born approximation at low energy nd temperature.

We propose to apply a quantum Monte Carlo technique based on Feynman-Kac 
formalism[5-7] of Quantum Mechanics which forms the basis for a simple and 
accurate way to calculate ground and excited state properties. To be precise, 
we apply Generalized Feynman Kac ( GFK) method [8-10] for attaractive and 
repulsive potentials for many boson system of alkali gases 
( Rb and Li respectively ).In our model the atoms in the dilute gas interact 
through Morse potential instead of conventional pseudopotential. Since at low 
temperature the de Broglie wavelength of the atoms become appreciable, we do 
full quantum treatment. To connect GFK to other many body techniques our 
numerical procedure ( GFK ) has a straightforward implementation to 
Schroedinger's wave mechanics. 
GFK is essentially a path integral technique with trial functions for which 
operations of the group of the wave function keep points in the chosen nodal 
region, provide an upper bound for the lowest state energy of that symmetry. 
The nodal region with the lowest energy serves as a least upper bound. If the 
nodal region has exact nodal structures of the true wave function the random 
walk is exact in the limit scale, time for walk, and number of walks get 
arbitrarily large. $Rb^{87}$ and $Li^{7}$ both having odd number of electrons 
and odd number of nuclei obey Bose statistics. So we do not need to 
worry about sign problem for these  calculations. Our method gives more 
accurate values for the ground and excited state energies as to calculate 
energy we approximate an exact solution ( i.e.,the GFK representation of it ) 
to the Schroedinger's equation, whereas most of the other numerical procedures 
approximate a solution to an approximate Schroedinger equation. 
Since we work with the bosonic system our results are exact within the 
numerical limitations.  
There are other Monte Carlo calculations which are worth mentioning in this 
context are ground state calculation by Diffision Monte Carlo method [11]and 
path integral Monte Carlo ( with temperature dependence )[12]. 
This paper is organized as follows. In Sec 2, we present the general formalism 
of Feynman-Kac and Generalized Feynman-Kac method and then describe it for 
the ground and excited state of trapped Bosons. In Sec 3, we describe the 
numeraical procedure. In Sec 4 we present results for positive ( $Rb^{87}$ ) 
and negative ( $Li^{7}$ ) scattering lengths. Finally in Sec 5 we summerize 
our achivements. 
\section{ Theory }
\subsection {  Path integral Theory at T=0}
\subsubsection { Feynman-Kac Path integretion }
For the Hamiltonian $H=-\Delta/2+V(x)$ consider the initial value problem
\ber
i\frac{\del u}{\del t}& =& (-\frac{\Delta}{2}+V)u(x,t)\nonumber\\
& &u(0,x)=f(x)
\eer
with $x \in  R^d$ and $u(0,x)=1$. The solution of the above equation can
be written in Feynman-Kac representation as
\bea
u(t,x)=E_xexp\{-\int_0^t V(X(s))ds\}
\eea
where X(t) is a Brownian motion trajectory and E is the average value of the
exponential term with respect to these trajectories. The lowest energy
eigenvalue for a  given symmetry can be obtained from the large deviation
priniciple of Donsker and Varadhan [13],
\bea
\lambda=-\lim_{t\rightarrow \infty} \frac{1}{t}ln E_x
exp\{-\int_0^t V(X(s))ds\}
\eea
Generalizations of the class of potential functions for which Eqs. 2 and 3 
are valid are given by Simon[14] and include most physically interesting 
potentials, positive or  negative, including, in particular, potentials
with $1/x$ singularities. It also means that the functions determined
by Eq(2) will be the one with lowest energy of all possible functions
independent of symmetry. Restrictions on allowed Brownian motions 
must be imposed to get a solution of the desired symmetry if it is not the 
lowest energy solution for a given Hamiltonian. Although other interpretations 
are interesting, the simplest is that the Brownian motion distribution
is just a useful mathematical construction which allows one  to extract
a physically relevant quantity, the ground and excited state energy of a 
quantum mechanical system. In numerical implementation of Eq(3) the 3N 
dimensional Brownian motion is replaced by 3N independent, properly scaled one 
dimensional random walks as follows. For a given time t and integers n and l
define [6] the vector in $R^{3N}$ 
\ber
W(l)\equiv W(t,n,l)
& = & ({w_1}^1(t,n,l),{w_2}^1(t,n,l),{w_3}^1(t,n,l)....\\ \nonumber
&   &                  .......{w_1}^N(t,n,l){w_2}^N(t,n,l){w_3}^N(t,n,l)
\eer  
where 
\bea
{w_j}^i(t,n,l)=\sum^l_{k=1}\frac{{\epsilon}^i_{jk}}{\sqrt n}
\eea
with ${w_j}^i(0,n,l)=0$ 
for $i=1,2,....N$;$j=1,2,3$ and $l=1,2,,,,,,nt$. Here $\epsilon $ is 
chosen independently and randomly with probability P for all i,j,k such that
$P({\epsilon}^i_{jk}=1)$=$P({\epsilon}^i_{jk}=-1)$=$\frac{1}{2}$
It is known by an invariance principle[15] that for every $\nu$ and W(l)
defined in Eq(4) 

\ber
\lim_{n\to\infty}P(\frac{1}{n}\sum^{nt}_{l=1}V(W(l)))\leq \nu \\ \nonumber
 =  P( \int\limits^t_0 V( X(s))ds\leq\nu
\eer
Consequently for large n,
\ber
P[ \exp(- \int\limits^t_0 V(X(s))ds)\leq\nu ] \\ \nonumber 
 \approx  P [\exp(-\frac{1}{n}\sum^{nt}_{l=1}V(W(l)))\leq \nu]
\eer
By generating $N_{rep}$ independent replications $Z_1$,$Z_2$,....$Z_{N_{rep}}$  of
\bea
Z_m=\exp(-(-\frac{1}{n}\sum^{nt}_{l=1}V(W(l)))
\eea
and using the law of large numbers, $(Z_1+Z_2+...Z_{N_{rep}})/N_{rep}=Z(t)$
is an approximation to Eq(2)
\bea
\lambda\approx -\frac{1}{t}logZ(t)
\eea
Here $W^m(l), m=1,2,N_{rep}$ denotes the $m^{th}$ realization of W(l) out of 
$N_{rep}$ independently run simulations. In the limit of large t and $N_{rep}$
this approximation approaches an equality, and forms the basis of a 
computational scheme for the lowest energy of a many particle system with
a prescribed symmetry.

In dimensions higher than 2, the trajectory X(t) escapes to infinity with
probability 1. As a result, the important regions of the potential are
sampled less and less frequently and the above equation converges slowly.
Now to speed up the convergence we use Generalized Feynman-Kac (GFK) method.
\subsubsection { Generalized Feynman Kac path integretion }
To formulate the (GFK) method, we first rewrite the
Hamiltonian as $H=H_0+V_p$, where
$ H_0=-{\Delta}/2+{\lambda}_T+{{\Delta}{\psi}_T}/{2{\psi}_T} $ and
$V_p=V-({\lambda}_T+\Delta{\psi}_T/2{\psi}_T)$.
Here ${\psi}_T$ is a twice differentiable nonnegative reference function
and $H{\psi}_T={\lambda}_T{\psi}_T$. The expression for the energy can now be
written as
\bea
\lambda={\lambda}_T-\lim_{t\rightarrow \infty}
 \frac{1}{t}ln E_xexp\{-\int_0^t V_p(Y(s))ds\}
\eea
where Y(t) is the diffusion process which solves the stochastic differential
equation
\bea
dY(t)=\frac{{\Delta}{\psi}_T(Y(t))}{{\psi}_T(Y(t))}dt+dX(t)
\eea
The presence of both drift and diffusion terms in this expression
enables the trajectory Y(t) to be highly localized. As a result, the important
regions of the potential are frequently sampled and Eq (3) converges rapidly.\\

\subsection{ Schroedinger Formalism for Rb and Li condensate }
Case 1: A system of few Rb atoms interacting through a model potential in an 
isotropic trap.\\
                                                                 
\bea
[-{\Delta}/2+V_{int}+\frac{1}{2} \sum_{i=1}^N [{x_i}^2
 +  {y_i}^2+{z_i}^2]\psi(\vec{r})
=E\psi(\vec{r})
\eea
The atoms here interact through a model potential
$Dsech^2(r/r_0)$\\
Case 2 : Systems of Rb and Li atoms interacting through Morse potential
trapped inside an anisotropic trap. We choose to work in the cylindrical 
coordinates as  the original experiment 
had an axial symmetry, cylindrical coordinates are the natural choices for this
problem. We consider a cloud of N atoms interacting through repulsive potential
placed in a three dimensional anharmonic oscillator potential. 
At low energy the motion of condensate can be represented as 
\bea
[-{\Delta}/2+V_{int}+\frac{1}{2} \sum_{i=1}^N [{x_i}^2
 +  {y_i}^2+\lambda{z_i}^2]\psi(\vec{r}) 
=E\psi(\vec{r}) 
\eea
where $\frac{1}{2} \sum_{i=1}^N [{x_i}^2
 +  {y_i}^2+\lambda{z_i}^2]\psi(\vec{r}) $ is the anisotropic 
potential with anisotropy factor $\lambda=\frac{{\omega}_z}{{\omega}_x}$.
Now 
 \bea
 V_{int}=V_{Morse}=
\sum_{i,j} V(r_{ij})=\sum_{i<j}D[e^{-\alpha(r-r_0)}(e^{-\alpha(r-r_0)}-2)]
 \eea
The above Hamiltonian is not separable in spherical polar coordinates because
of the anisotropy. In cylindrical coordinates the noninteracting part behaves 
as a system of noninteracting harmonic oscillators and can be writtem as 
follows :
\ber
& & [-\frac{1}{2 \rho}\frac{\partial}{\partial \rho}(\rho\frac{\partial}
{\partial \rho}) -\frac{1}{{\rho}^2}\frac{{\partial}^2}{\partial {\phi}^2}
-\frac{1}{2}\frac{{\partial}^2}{\partial{z}^2}\nonumber \\
& & +\frac{1}{2}({\rho}^2+{\lambda}^2z^2)]\psi(\rho,z) \nonumber \\
& = & E \psi(\rho,z)
\eer
The  energy 'E' of the above equation can be calculated exactly which is
\bea
 E_{n_{\rho} {n_z} m}
=(2n_{\rho}+|m|+1)+(n_z+1/2)\lambda
\eea

Since we are dealing with many Boson systems having very small
number of  particles, ( In JILA experiment the number
of particles is of the order of 2000 and does not correspond to Thomas-Fermi
limit ). So it is quite legitimate to use Gaussian trial functions for 
modes which are not coupled.
In our guided random walk we use the  solution of
Schoroedinger equation for harmonic oscillator in d-dimension 
as the trial function as follows [16]:
 
\bea
{\psi}_{n_{\rho}{n_z}m}(\vec r)\simeq \exp^{\frac{-z^2}{2}}H_{n_z}(z)\times
e^{im\phi}{\rho}^me^{-{{\rho}^2}/2}{L_{n_{\rho}}}^{(m)}({\rho}^2)
\eea
\subsection{  Fundamentals of BEC }                                            
Even though the phase of Rb vapors at T=0 is certainly solid, Bose condensates 
are preferred in the gasous form over the liquids and solids because 
at those higher densities interactions are complicated and hard to deal with 
on an elementary level. They are kept metastable by maintaining a very low 
density. For alkali metals, $\eta$, the ratio of zero point energy and 
molecular binding energy lies between $10^{-5}$ and $10^{-3}$. According
to the theory of corresponding states[17] since for the T=0 state of alkali 
metals, $\eta$ exceeds a critical value 0.46, the  molecular binding energy 
dominates over the zero point motion and they condense to solid phase. 
 But again the life time of a gas is limited by three body recombination rate 
which is proportional to the sqaure of the atomic density. 
It gets suppressed at low density. Magnetically trapped alkali vapors can be 
metastable depending on their densities and lifetimes. 
So keeping the density low only two body collisions are allowed as a result of 
which dilute gas approximation [11] still holds for condensates which 
tantamounts to saying $na^3<<1$ (a is the scattering length of s wave).
Now defining $n=N/V=r_{av}^{-3}$ as a mean distance between the atoms 
( definition valid for any temperature ), the dilute gas condition reads as
$a<<r_av$ and zero point energy dominates (dilute limit). In the dense limit,
for $a \approx r_{av} $ on the other hand the interatomic potential dominates
[11].The gas phase is accomplished by reducing the material density through 
evaporative cooling.
\section{ Numerical procedure }
\subsection{ Dilute limit }
In the dilute limit and at very low energy only binary collisions are possible
and no three body recombination is allowed. In such two body scattering at
low energy first order Born approximation is applicable and the interaction
strength 'D' can be related to the single tunable parameter of this problem,
the s-wave scattering length 'a' through the relation given below. This single
parameter can specify the interaction completely without the detail of the
potential in the case of pseudopotentials. We study a system of Rb particles 
interacting through following two repulsive potentials. \\
A. Model potential $ V(r)= Dsech^2(r/r_0) $\\
B. Morse potential for dimer of rubidium \\
 \bea
\sum_{i,j} V(r_{ij})=\sum_{i<j}D[e^{-\alpha(r-r_0)}(e^{-\alpha(r-r_0)}-2)]
 \eea
with more realistic feature of having repulsive core at
$r_{ij}=0$ than other model potentials. In our case, the interaction strength
depends on two more additional parameters, $ r_0 $ and $ \alpha $.
\bea
a=\frac {mD}{4 \pi {\hbar}^2}\int V(r)d^3r
\eea
Case I. For potential 'A', the above yields 
$ D= \frac{12 a {\hbar}^2}{m {\pi}^2{\vec{r_0}}^3}$
The Hamiltonian for a system of Rb gas in an isotropic trap intertacting
throuh a potential $V(r)$ can be written as
\ber
& &[ -{\hbar}^2/2m
\sum_{i=1}^N {\nabla^{\prime}_i}^2
+ \sum_{i,j}V(r^{\prime}_{ij}) \nonumber \\
& &  +\sum_{i=1}^N\frac{m}{2} {\omega_i}^2({x^{\prime}_i}^2
+ {y^{\prime}_i}^2+{z^{\prime}_i}^2)]\psi(\vec{r}^{\prime}) \nonumber \\
& = &E\psi(\vec{r}^{\prime})
\eer
The above Hamiltonian can be rescaled by substituting
$\vec{r}^{\prime}=s\vec{r}$ and $E=E_0U$ and the interaction strengh
in the dimensionless form can be written as
\bea
{\gamma}_1=\frac{12 a }{s {r_0}^3{\pi}^2}
= 3.05 \times 10^{-3}
\eea
with $r_0=1.2$.\\
Case II. Again for  potential 'B' the Born approximation yields
\bea
 D=\frac{4{\hbar}^2a{\alpha}^3}{m e^{\alpha r_0}(e^{\alpha r_0}-16)}
\eea
The Hamiltonian for Rb gas with an asymmetric trapping potential and
Morse type mutual interaction can be written as
\ber
& &[ -{\hbar}^2/2m
\sum_{i=1}^N {\nabla^{\prime}_i}^2
+ \sum_{i,j}V(r^{\prime}_{ij}) \nonumber \\
& &  +\frac{m}{2}({\omega_x}^2\sum_{i=1}^N {x^{\prime}_i}^2
+{\omega_y}^2\sum_{i=1}^N {y^{\prime}_i}^2+{\omega_z}^2\sum_{i=1}^N
{z^{\prime}_i}^2)]\psi(\vec{r}^{\prime}) \nonumber \\
& = &E\psi(\vec{r}^{\prime})
\eer
The above Hamiltonian can be rescaled by substituting
$\vec{r}^{\prime}=s\vec{r}$ and $E=E_0U$ as
\ber
& &[ -\frac{{\hbar}^2}{2ms^2}
\sum_{i=1}^N {\nabla_i}^2
 +  \sum_{i<j}\frac{4{\hbar}^2a{\alpha}^3}{m s^3
e^{\alpha r_0}(e^{\alpha r_0}-16)}[e^{-\alpha(\vec{r_{ij}}-r_0)}(e^{-\alpha
(\vec{r_{ij}}-r_0)}-2)] \nonumber \\
& & + \frac{ms^2}{2}({\omega_x}^2\sum_{i=1}^N {x_i}^2
 + {\omega_y}^2\sum_{i=1}^N {y_i}^2+
{\omega_z}^2\sum_{i=1}^N
{z_i}^2)]\psi(\vec{r}) \nonumber \\
& = & E_0 U\psi(\vec{r})
\eer
\ber
& & [ \frac{1}{2}\sum_{i=1}^N {\nabla_i}^2
-4\frac{a{\alpha}^3}{se^{\alpha r_0}(e^{\alpha r_0}-16)}\sum_{i<j}
[e^{-\alpha(r_{ij}-r_0)}(e^{-\alpha(r_{ij}-r_0)}-2)] \nonumber \\
& & -\frac{m^2{\omega_x}^2s^4}{2\hbar^2}
\sum_{i=1}^N ({x_i}^2
 + \frac{{\omega_y}^2}{{\omega_x}^2} {y_i}^2+
\frac{{\omega_z}^2}{{\omega_x}^2}{z_i}^2)]\psi(\vec{r})\nonumber \\
 & = & -E_0 \frac{Ums^2}{\hbar^2}\psi(\vec{r})
\eer
Let $\frac{m^2{\omega_x}^2s^4}{\hbar^2}=1\Rightarrow
s^2= \frac{\hbar}{m\omega_x}$ is the natural unit of length. Let
$\frac{Ums^2}{\hbar^2}=1\Rightarrow U=\frac{\hbar^2}{ms^2}=\hbar\omega_x$
is the natural unit of energy.
Then the standard form of the equation  becomes
\ber
& & [\frac{1}{2}\sum_{i=1}^N {\nabla_i}^2
-\sum_{i<j}4\frac{a{\alpha}^3}{se^{\alpha r_0}(e^{\alpha r_0}-16)}\sum_{i<j}
[e^{-\alpha(r_{ij}-r_0)}(e^{-\alpha(r_{ij}-r_0)}-2)] \nonumber \\
& & -\frac{1}{2} \sum_{i=1}^N ({x_i}^2
 + \frac{{\omega_y}^2}{{\omega_x}^2} {y_i}^2+
\frac{{\omega_z}^2}{{\omega_x}^2}{z_i}^2)]\psi(\vec{r}) \nonumber\\
& = & - E_0 \psi(\vec{r})
\eer
With   $\omega_x=\omega_y=\frac{\omega_z}{\sqrt \lambda}$, the above eqn
becomes,
\ber
& & [\frac{1}{2}\sum_{i=1}^N {\nabla_i}^2
-4\frac{a{\alpha}^3}{se^{\alpha r_0}(e^{\alpha r_0}-16)}\sum_{i<j}
[e^{-\alpha(r_{ij}-r_0)}(e^{-\alpha(r_{ij}-r_0)}-2)] \nonumber \\
& & -\frac{1}{2} \sum_{i=1}^N [{x_i}^2
 +  {y_i}^2+\lambda{z_i}^2]\psi(\vec{r}) \nonumber\\
& = & - E_0 \psi(\vec{r})
\eer
\ber
& & [\frac{1}{2}\sum_{i=1}^N {\nabla_i}^2
-\gamma\sum_{i<j}
[e^{-\alpha(r_{ij}-r_0)}(e^{-\alpha(r_{ij}-r_0)}-2)] \nonumber \\
& & -\frac{1}{2} \sum_{i=1}^N [{x_i}^2
 +  {y_i}^2+\lambda{z_i}^2]\psi(\vec{r}) \nonumber\\
& = & - E_0 \psi(\vec{r})
\eer
Now for $\alpha=.35$ and $r_0=11.65$ (both in oscillator units)[4,18], the
Morse potetial become almost noninteracting and the results become 
substantially lower than GP. We find a better agreement with GP 
by choosing  $\alpha=.29$ and $r_0=9.67$. 
We have checked that for these choice of parameters, Morse solution is
extremely good. $ a=52 \times 10^{-10}$, $s=.12\times 10^{-5}$, the interaction
strength $\gamma$ is given by
\bea
\gamma_2=4\frac{a{\alpha}^3}{se^{\alpha r_0}(e^{\alpha r_0}-16)}
=4.9\times 10^{-5}
\eea
For mean field calculation the value of interaction strength was taken to be
$ 4.33 \times 10^{-3}$. For this problem we are interested 
in the limit $\gamma << 1 $. The case $\gamma >> 1 $ is usually known as the
Thomas Fermi limit. For  $ \gamma= 4.9 \times 10^{-5}$, the
eigenvalue equation reduces to a minimally perturbed system of d dimensional
anisotropic oscillator where $d=3N $ and N is the number of particles.
The whole concept of bound states of Morse dimers is very outside the range of
this limit, so the nonexistence of two-body bound states is ensured by 
choosing the above parameters.

Case III. Using the same Morse potential with $\alpha=0.4$ and $r_0=6.8$,  
one can generate  attractive interaction for $Li^{7}$ as well.  In this case 
the interaction strength turns out to be 
\bea
\gamma_3=4\frac{a{\alpha}^3}{se^{\alpha r_0}(e^{\alpha r_0}-16)}
=-9.345\times 10^{-6}
\eea
Even though  $\gamma << 1 $,  we solve the
eigenvalue eqn nonperturbatively with Generalized Feynman-Kac  procedure.
Energies  at zero temperature are obtained for ground and excited states by 
solving Eq.(1) and using Eq.(3).  Since original Feynman-Kac method [5-6] is 
computationally inefficient we incorporate  importance sampling in our random 
walk and use trial function of the form given in Eq.(17)
                                                                                

\newpage
\section{ Results}
Case 1. A system of Rb atoms in an isotropic trapping and interactinng
through a  model potential \\ 
$ D sech^2(r/r_0)$.
\begin{table}[h!]
\caption{In column 2, ground state energies for 3-20 particles in an 
isotropic trap with potential 'a' in units of $\hbar \omega_{\perp} $ for 
$a=.00305 a_{ho}$. Columns 3-5 represent the same energies by 
Blume and Green(BG)[19], GP and modified GP respectively}.
\begin{center}
\begin{tabular}{ccccc} \hline\hline
N  & $ E_{GFK} $ & $E_{BG} $ & $E_{GP}$ & $ E_{GP,mod}$ \\ 
\hline
3 &   4.50925(3) & 4.51037(2) & 4.51032 & 4.51032 \\
5 &   7.53086(5) & 7.53439(6) & 7.53432 & 7.53434 \\
10&   15.13867(3)& 15.1539(2) & 15.1534 & 15.1535 \\
20 &  30.58460(4)& 30.639(1)  & 30.638  & 30.639  \\
\hline\hline
\end{tabular}
\end{center}
\end{table}
\subsection{\bf Positive scattering length: Rb}
As the potential does not sustain any many body bound state ( which is ensured
by suitably choosing the value of the parameters $\alpha$ and $r_0$ ) and the
scattering length is positive the  system behaves as a gas or as a
metastable state which can be long-lived at very low densities[3].
In the table below, we explicitly show the expectation values of trap 
potential, interatomic potential and kinetic energy as three components of 
total energy for different number of particles and it is observed that virial 
theorem is satisfied in each case[20].
\bea
2E_{kin}-2E_{HO}+3E_{pot}=0
\eea
From Fig 1, 2 and 3, we see that energy/particle rises with increase in 
number of atoms in the trap for different symmetry states.
From Fig 4,5 and 6, we see that in the case of Rb,
increase in number of particles in the trap lowers the central density.
From the ground state data,the aspect ratio of velocity distribution 
i.e., $\sqrt{\frac{<p_z^2>}{<p_x^2>}}$ is calculated and its values are given 
in the last column of  Table 2. The value of aspect ratio in the noninteracting limit corresponds 
to $\sqrt \lambda$ and with increase in number of particles this value 
increases and in the Thomas Fermi limit it should correspond to $\lambda $. 
From our data, we verify this trend
of aspect ratio( Fig 7 ). 
\begin{table}[h!]
\caption{ Results for ground state of Rb with $\lambda=\sqrt 8 $ Chemical 
potential and energy are in units of $\hbar \omega_{\perp} $ and length is
in units of $a_{\perp}$. Numbers in the brackets correspond to the 
 reference[20]}
\begin{center}
\begin{tabular}{cccccccc} \hline\hline
N & $\mu$ & E/N & ${E/N}_{kin} $ & ${E/N}_{ho}$
& ${E/N}_{pot}$ & $\sqrt {< x^2>/<z^2>} $\\ \hline
1  & 2.414213 &2.414213  & 1.207409  & 1.206803  &  0.0      &  1.679563    \\
   &          &          &           &          &           &  (1.68333)\\ 
10 & 2.448952& 2.431595(5) & 1.202488   & 1.211725  &.017369  & 1.684900     \\
40 & 2.564350 &2.489287(1)  &1.196455   &1.217758  & 0.075068  &  1.68732   \\
70 & 2.678893 &2.546602(6)  &1.188591   & 1.225621  & 0.132339   &1.688079      \\
100 & 2.792482 &2.603549(3)   &1.180656  & 1.233556 & 0.189134   &1.688960     \\
   & (2.88)   &   (2.66) &           &          &           &  (1.79545)         \\
200  & 3.149535  &  2.79075(7)  & 1.061734  & 1.35247& 0.367660   &1.690056\\
   & (3.21)   & (2.86)   &           &          &           &  (1.88888)
 \\  
\hline\hline
\end{tabular}
\end{center}
\end{table}
\newpage
\begin{figure}[h!]
\centering
\epsfxsize=4in{\epsfbox{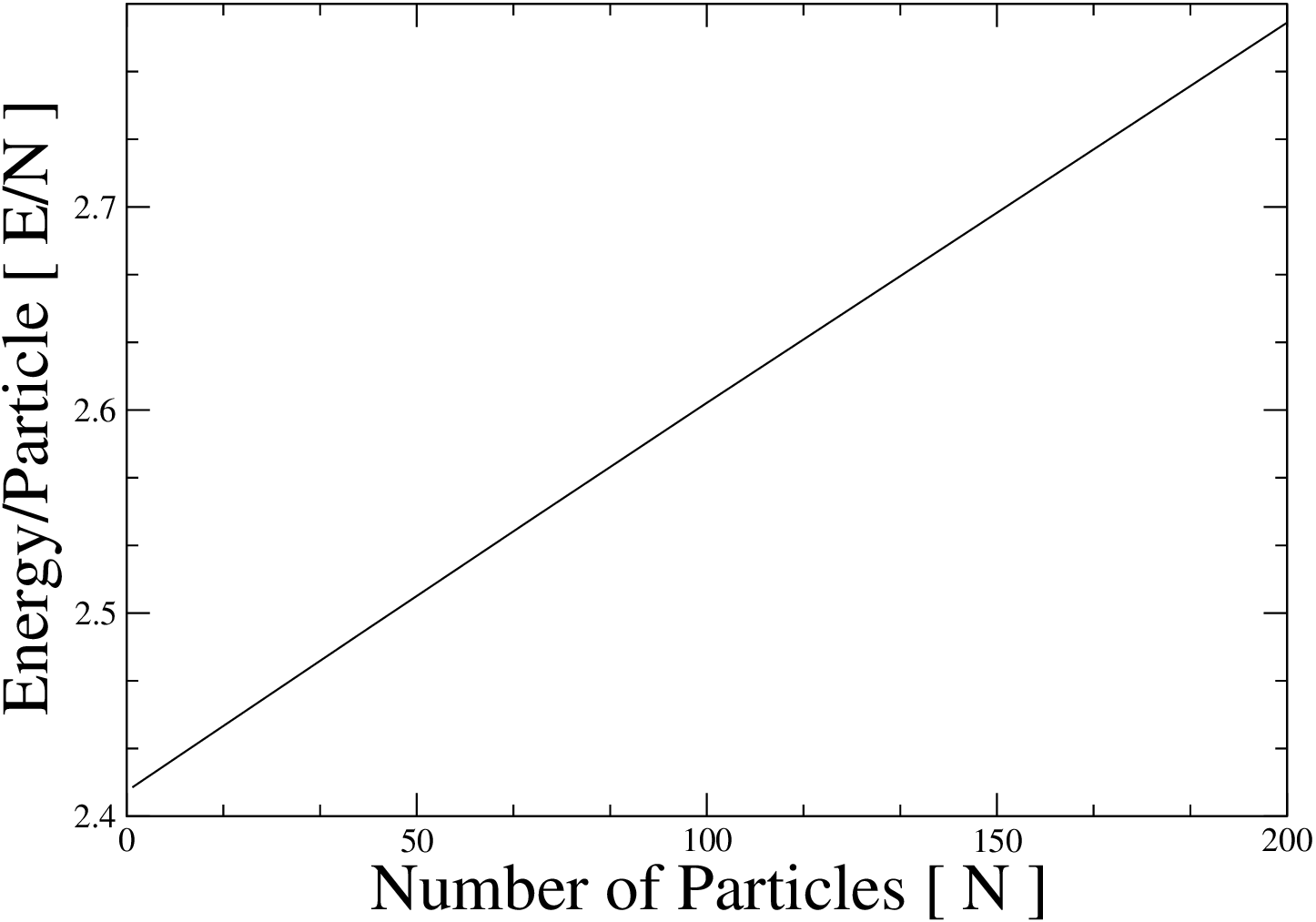}}
\caption{A plot for the Condensate Energy/Particle  versus Number of atoms in
trap for 200 particles for the ground state :
$n_z=n_{\rho}=m=0$; this work}
\end{figure}

\vskip 2cm
\begin{table}[h!]
\caption{ Results for excited states of Rb gas with  $\lambda=\sqrt 8 $ }
\begin{center}
\begin{tabular}{ccccc} \hline\hline
N & ${E/N}_{n_z=n_{\rho}=0,m=1}$ &${E/N}_{ n_z=n_{\rho}=0,m=2}$ & ${E/N}_{n_z=n_{\rho}=0,m=3}$ \\ \hline 
10 & 3.428526(1)  & 4.424937(2)               & 5.422566 (1)  \\
40 & 3.474993(3)  & 4.462418(3)               & 5.445864(4)   \\
70 & 3.520849(5)  & 4.500300 (4)              & 5.472183(6)  \\
100& 3.566892(4)  & 4.537793 (6)              & 5.494076(7)  \\  
200& 3.718571(8)  & 4.662026(7)               & 5.590799(8)   \\       
\hline
\end{tabular}
\end{center}
\end{table}
\newpage
\begin{figure}[h!]
\centering
\epsfxsize=3.5in{\epsfbox{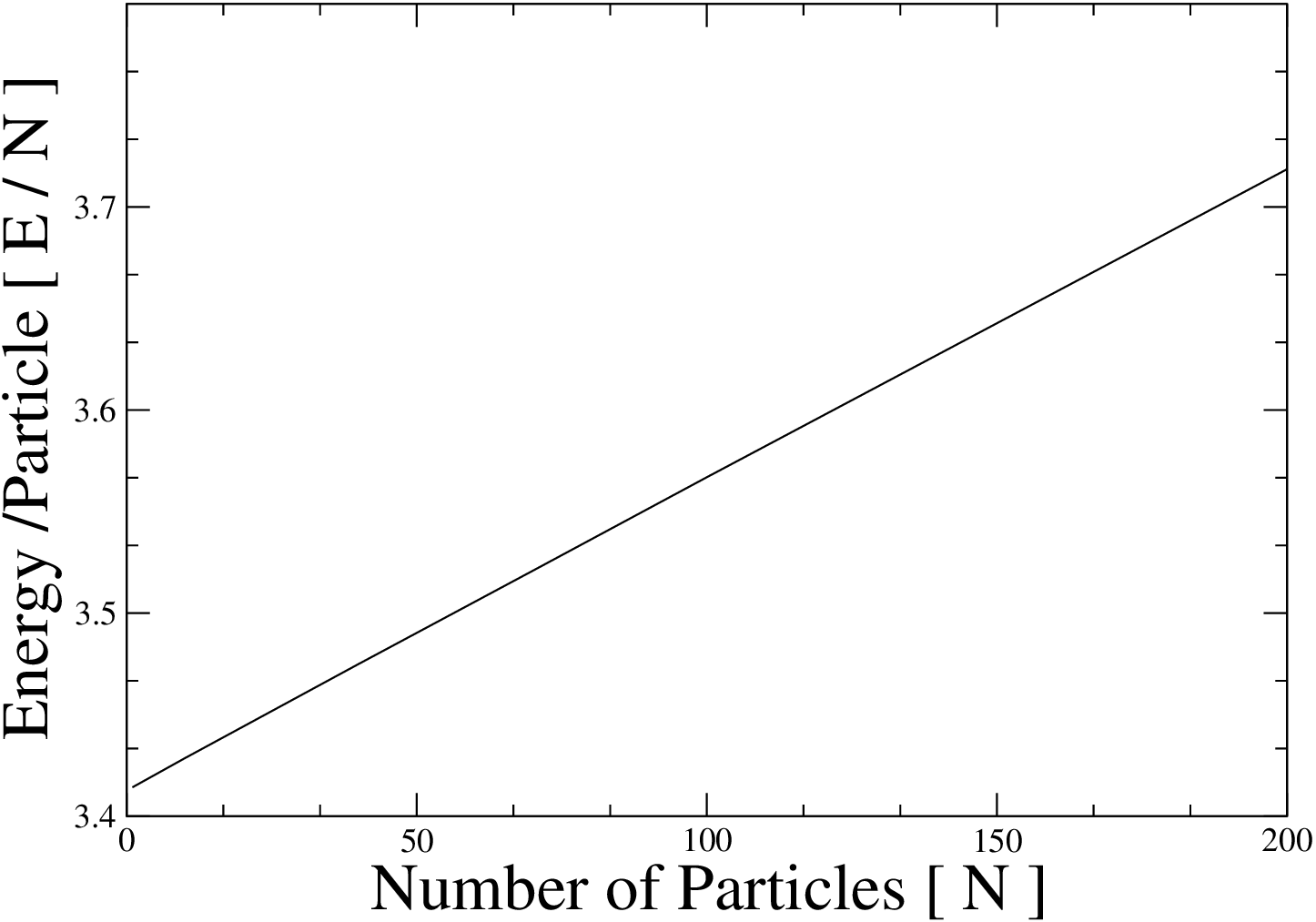}}
\caption{A plot for the Condensate Energy/Particle  versus Number of atoms in
trap for 200 particles for the 1st excited state : $n_z=n_{\rho}=0,m=1$;this
work}
\vskip 0.5cm
\end{figure}
\vskip 1cm
\begin{figure}[h!]
\centering
\epsfxsize=3.5in{\epsfbox{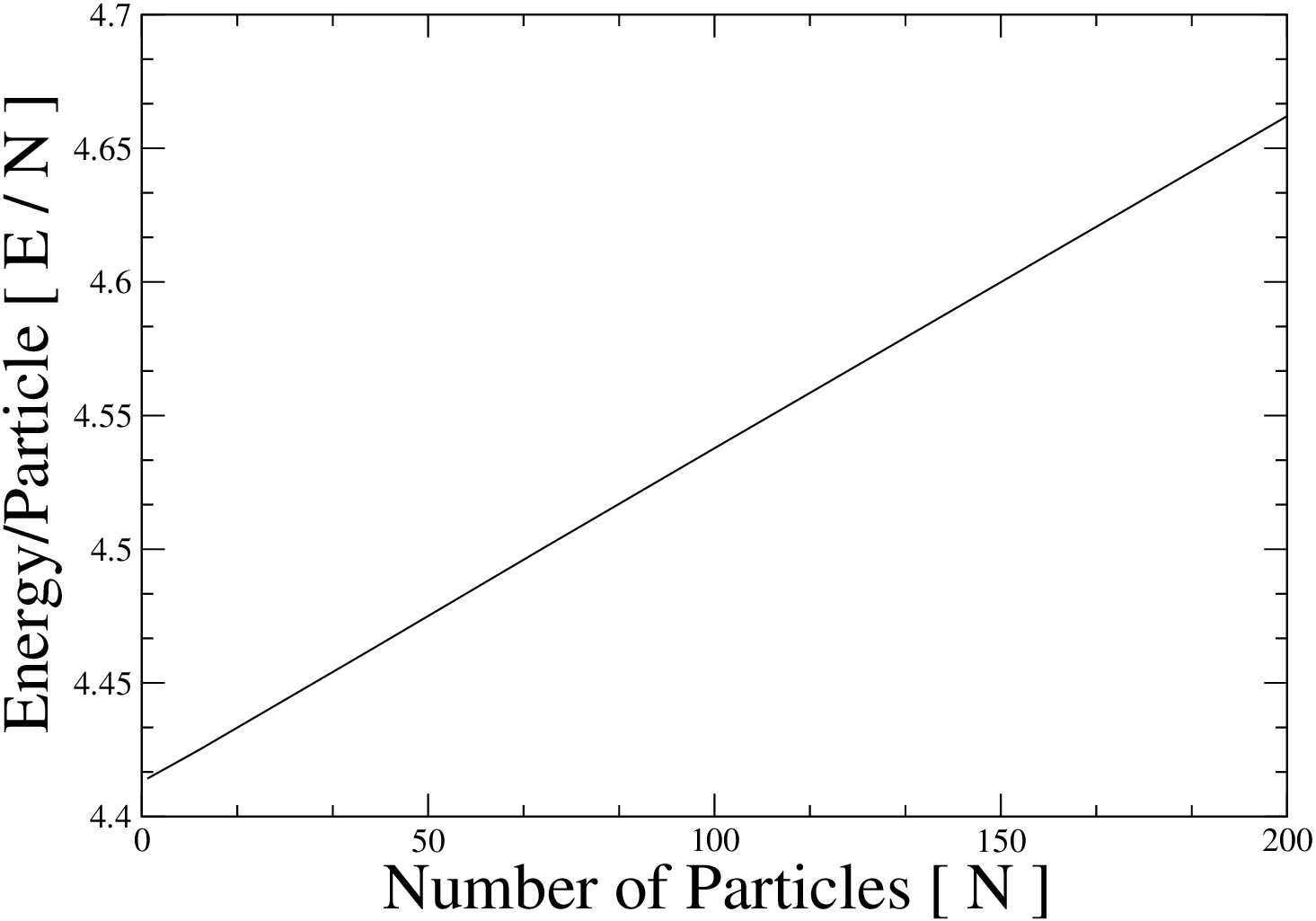}}
\caption{A plot for the Condensate Energy/Particle  versus Number of atoms in
trap for 200 particles for the 2nd excited state:
$n_z=n_{\rho}=0,m=2$; this work}
\vskip 2cm 
\end{figure}
\newpage
\begin{figure}[h!]
\centering
\epsfxsize=3.5in{\epsfbox{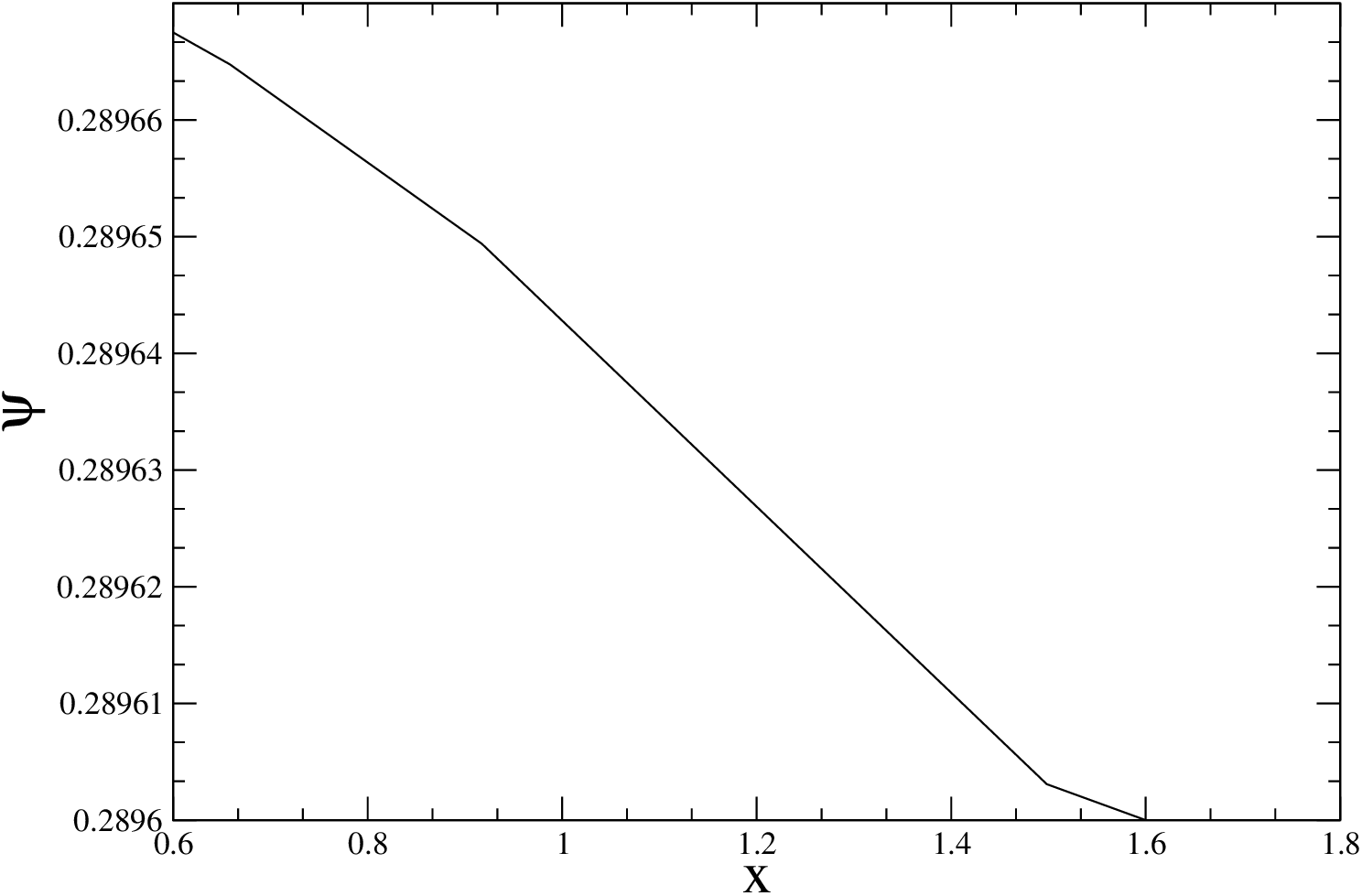}}
\caption{Ground state wave function of Rb along the x axis for N = 10
[this work] }
\end{figure}
\vskip 0.5cm
\begin{figure}[h!]
\centering
\epsfxsize=3.5in{\epsfbox{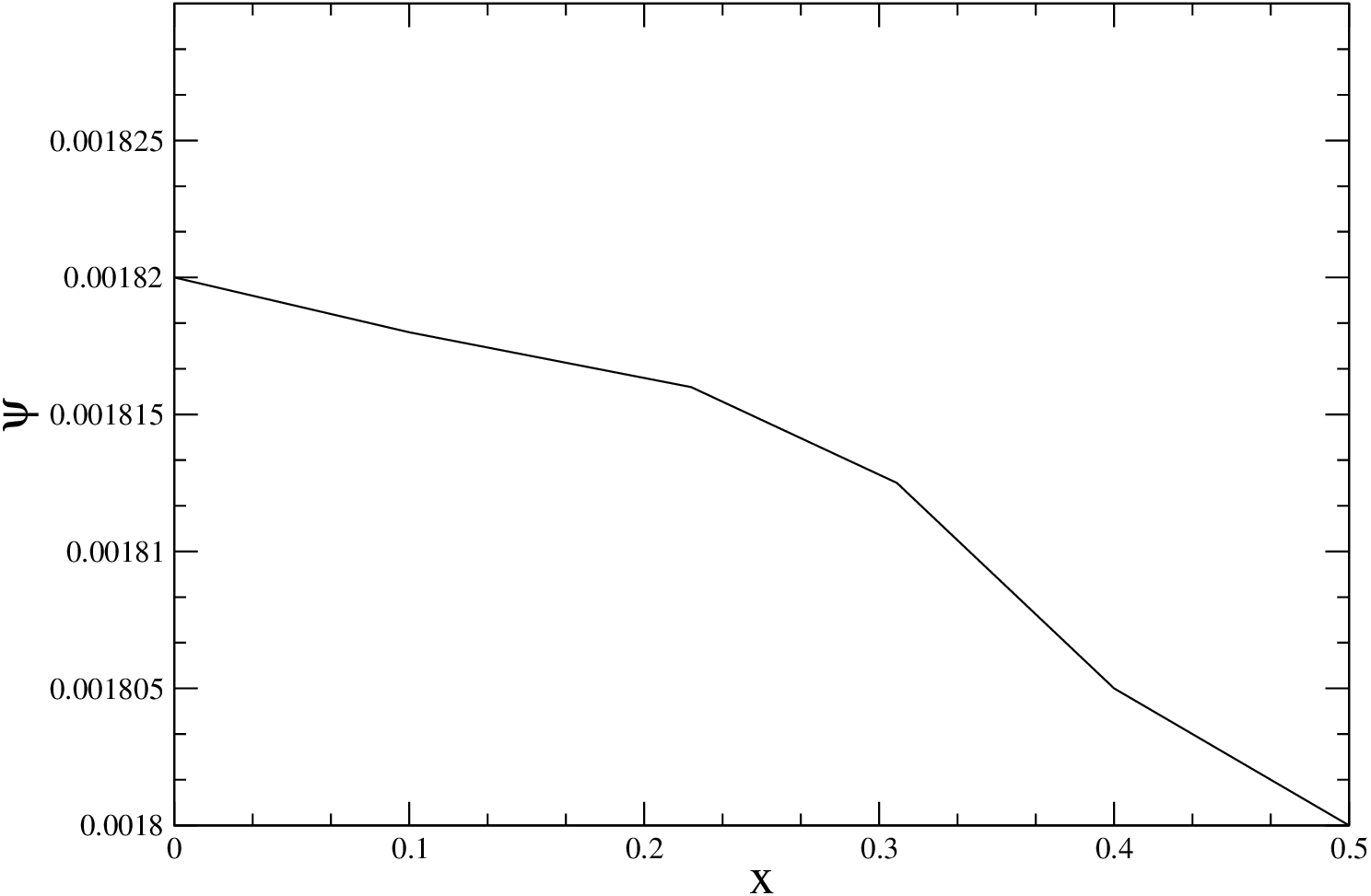}}
\caption{Ground state wave function of Rb along the x axis for N = 40
[this work] }
\end{figure}
\newpage
\begin{figure}[h!]
\centering
\epsfxsize=3.5in{\epsfbox{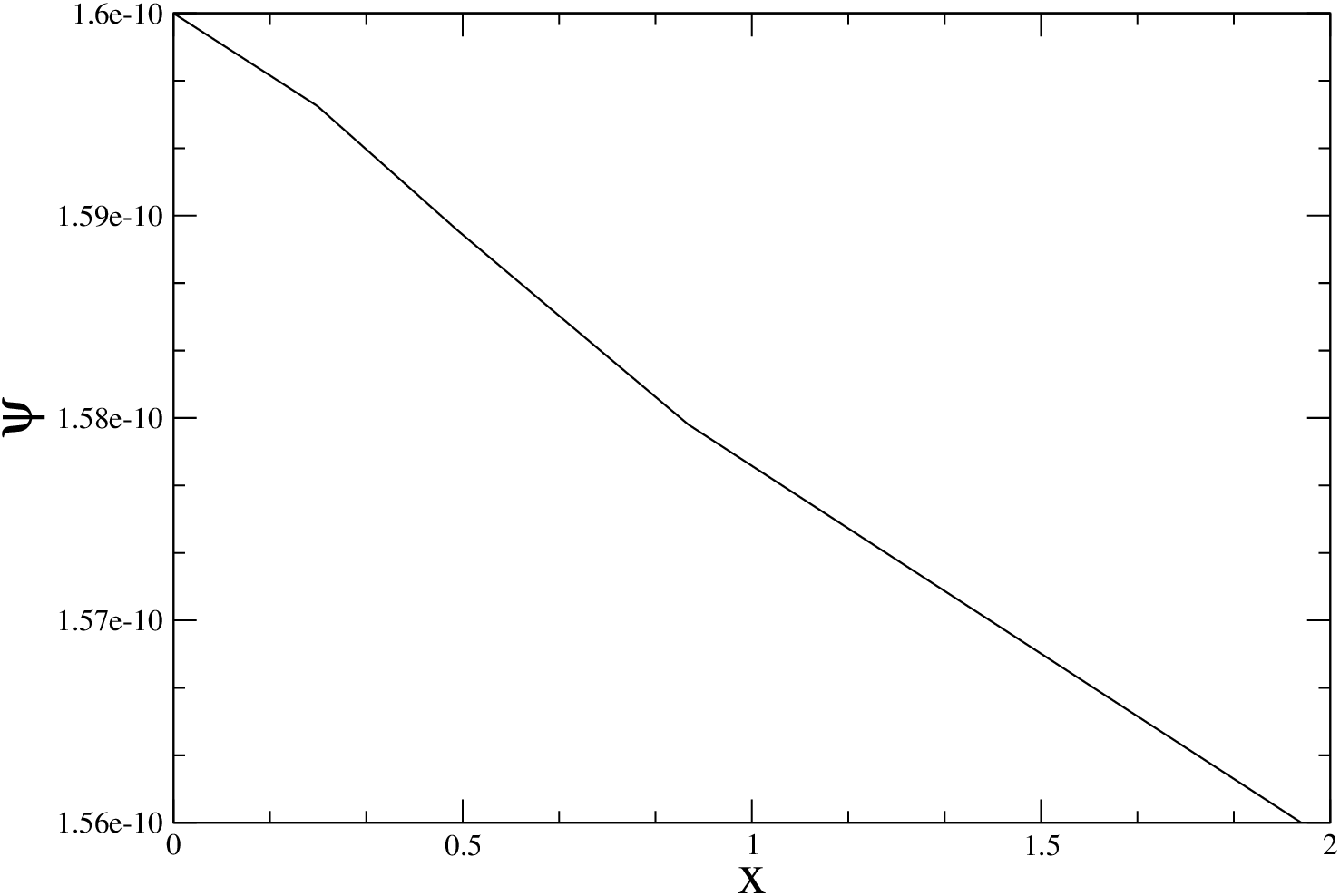}}
\caption{Ground state wave function of Rb along the x axis for N = 100
[this work] }
\end{figure}
\vskip 0.5cm
\begin{figure}[h!]
\centering
\epsfxsize=3.5in{\epsfbox{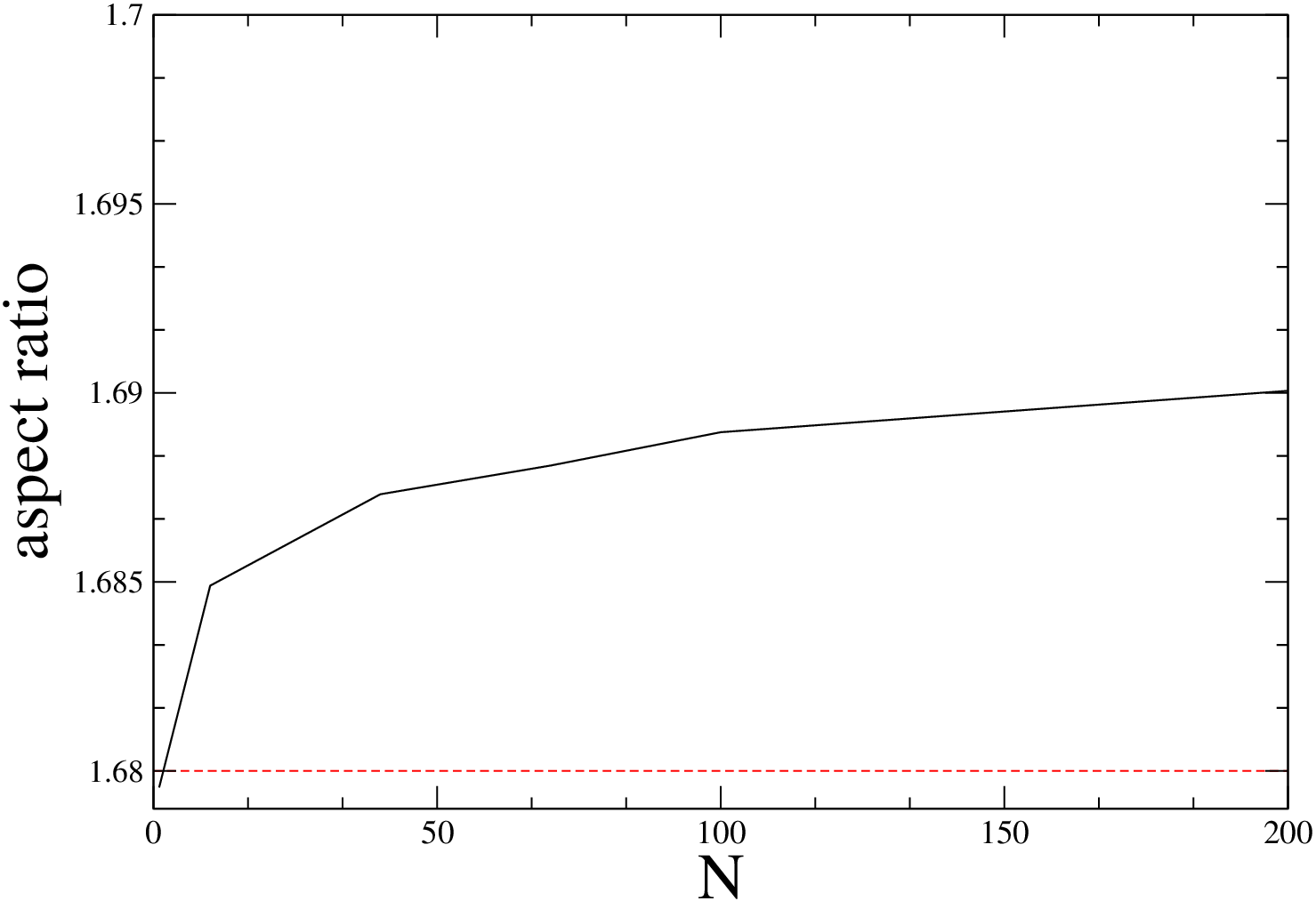}}
\caption{Aspect ratio in $Rb^87$ as a function of N. The horizontal line 
corresponds to a number close to $\sqrt \lambda $ [this work] }
\end{figure}
\newpage
\subsection{\bf Negative scattering length: Li}
In this section we report the calculations for Li. 
Since Li has negative scattering length, in Li gas the atoms interact with 
each other with attractive force and when this force becomes larger than
molar binding energies the gas collapses. But it is observed that if the 
number of particles in the gas is not too high, then zero point energy still
can exceed this attaractive force and still there can be a metastable state. 
Since we are dealing with only 200 atoms this does not pose any threat to our 
work. As Morse potential can be used both for positive and negative 
scattering lengths[21], we a use Morse potential for Li gas too and choose 
$\alpha = 0.4 $ and $ r_0 = 6.8 $ to produce an attractive type of interaction 
between Li atoms. In this case anisotropy factor is $\lambda =0.72 $ and 
scattering length is $-27 a_0 $. From Fig 8, 9 and 10, it is evident that 
increase in number of particles sharply increases the central density with 
contrast to Rb gas with positive scattering length where it decreases with 
increase in number of particles. In Fig 11, we plot aspect ratio of Li gas as 
a function of number of particles in the gas.
\newpage
\begin{table}[h!]
\caption{ Results for ground state of Li, $\lambda=\sqrt 8 $ Chemical
potential and energy are in units of $\hbar \omega_{\perp} $ and length is
in units of $a_{\perp}$. Numbers in the brackets correspond to the 
 reference[22]}
\begin{center}
\begin{tabular}{cccccccc} \hline\hline
N & $\mu$ & E/N & ${E/N}_{kin} $ & ${E/N}_{ho}$
& ${E/N}_{pot}$ & $\sqrt  {< x^2>/<z^2>} $\\ \hline
1  & 1.360000 &  1.360000 & 0.680273 &  0.679726  & 0.00000 &  0.847555   \\
   &          &  1.360000 &          &            &         &  (0.84873)   \\  
10 & 1.354930 &  1.357467(1) & 0.680486 &  0.6795132 &-0.002534 &  0.848088  \\ 
40 & 1.338018 &  1.349014(3) & 0.681689 &  0.678310  & -0.010990 & 0.848529\\ 
70 & 1.32108  &  1.340545(5) & 0.682509 &  0.677490  & -0.019458 & 0.848998  \\ 
100& 1.304113 &  1.332061(6) & 0.683446 &  0.676553  & -0.027943 & 0.849123 \\
   & (1.327)  &  (1.344)  &          &            &           & (0.850728)   \\
200& 1.247306 &  1.303674(7) & 0.687241 &  0.672758  & -0.056346 & 0.849731  \\
   & (1.291)  &  (1.326)  &          &            &           & (0.854858) \\ 
\hline\hline    
\end{tabular}
\end{center}
\end{table}
\begin{figure}[h!]
\centering
\epsfxsize=3.5in{\epsfbox{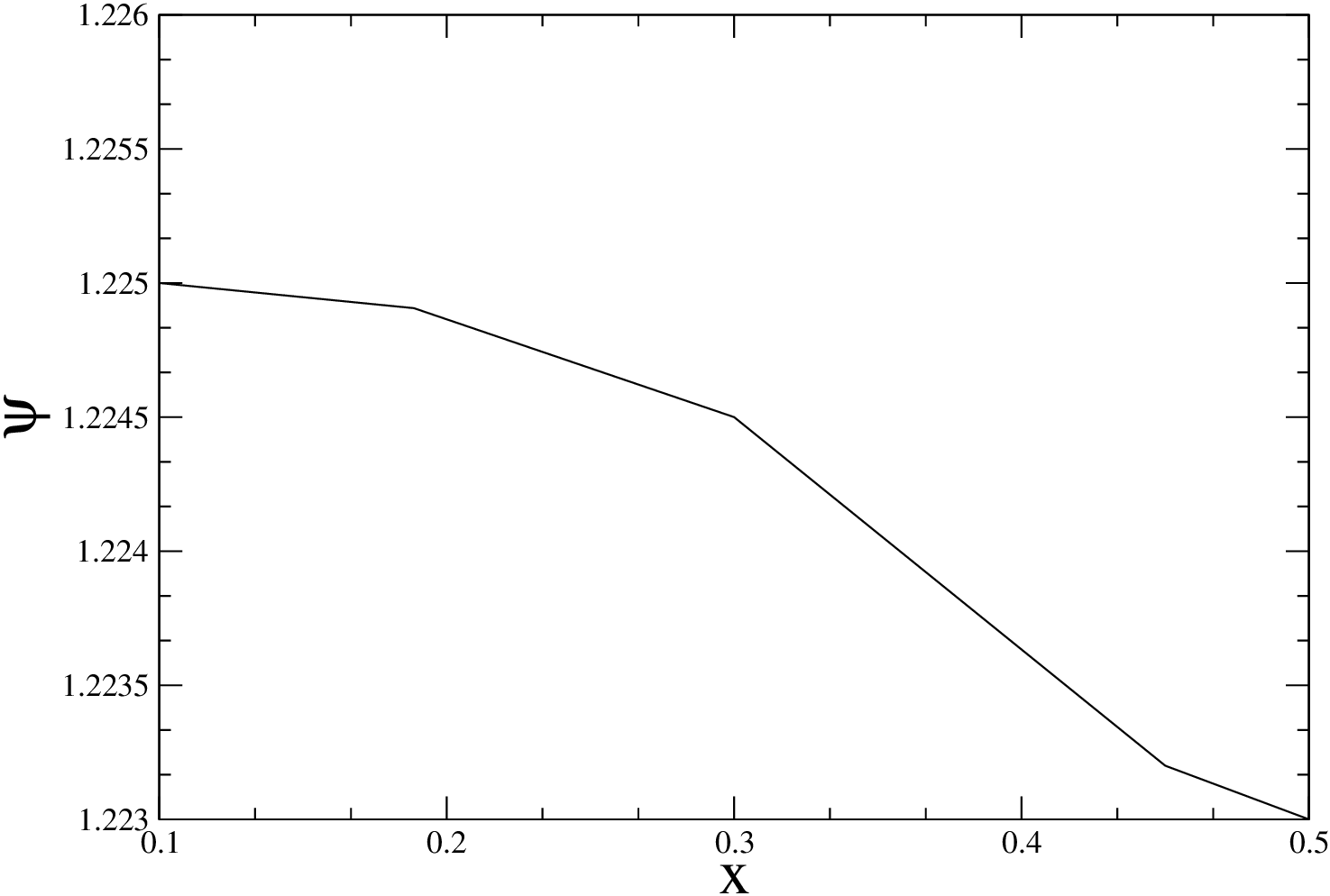}}
\caption{Ground state wave function of Li along the x axis for N = 10 
[this work] }
\end{figure}
\newpage
\begin{figure}[h!]
\centering
\epsfxsize=3.5in{\epsfbox{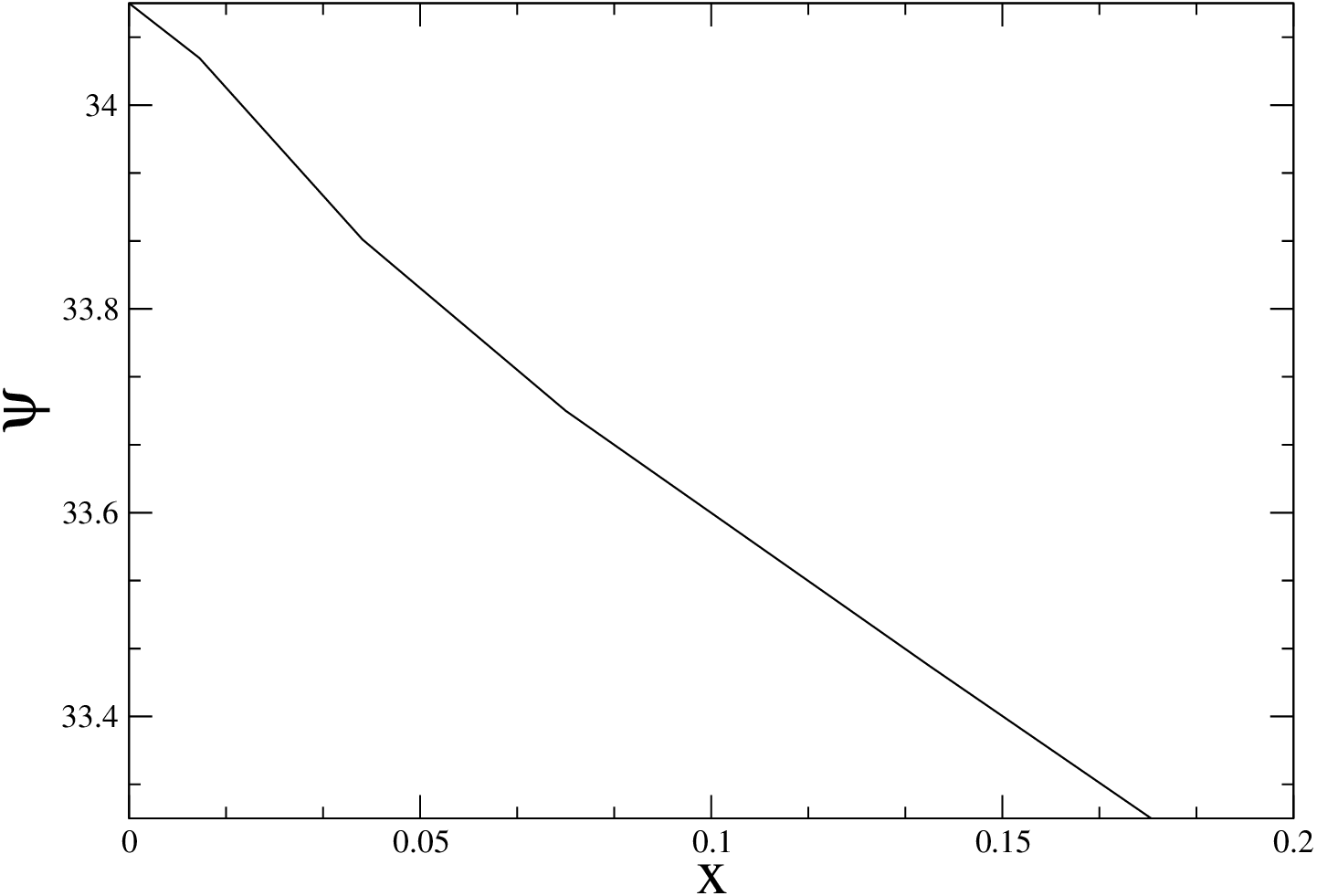}}
\caption{Ground state wave function of Li along the x axis for N = 40
[this work] }
\end{figure}
\vskip 0.5 cm
\begin{figure}[h!]
\centering
\epsfxsize=3.5in{\epsfbox{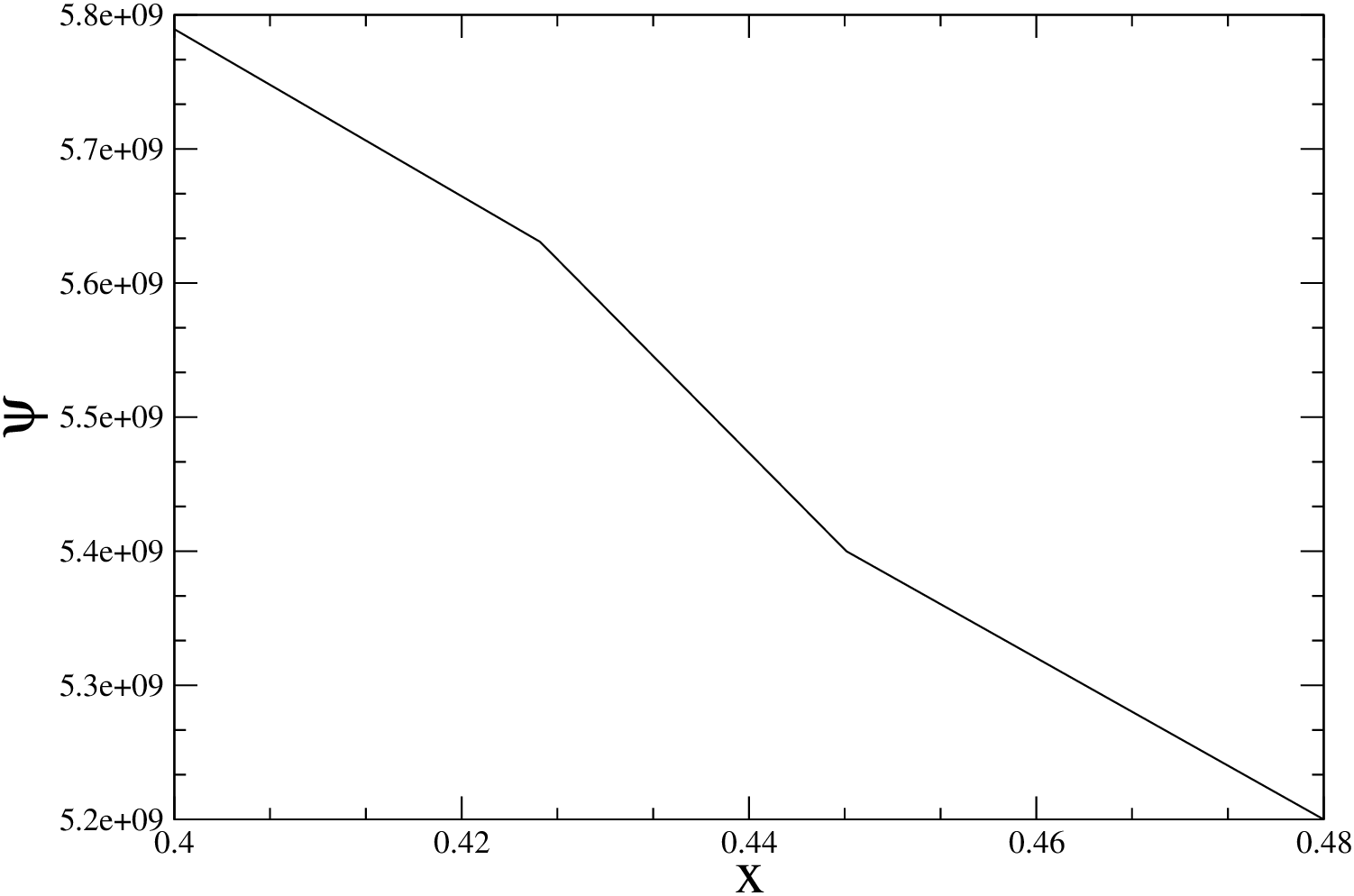}}
\caption{Ground state wave function of Li along the x axis for N = 100
[this work] }
\vskip 2cm
\end{figure}
\newpage
\begin{figure}[h!]
\centering
\epsfxsize=4.0in{\epsfbox{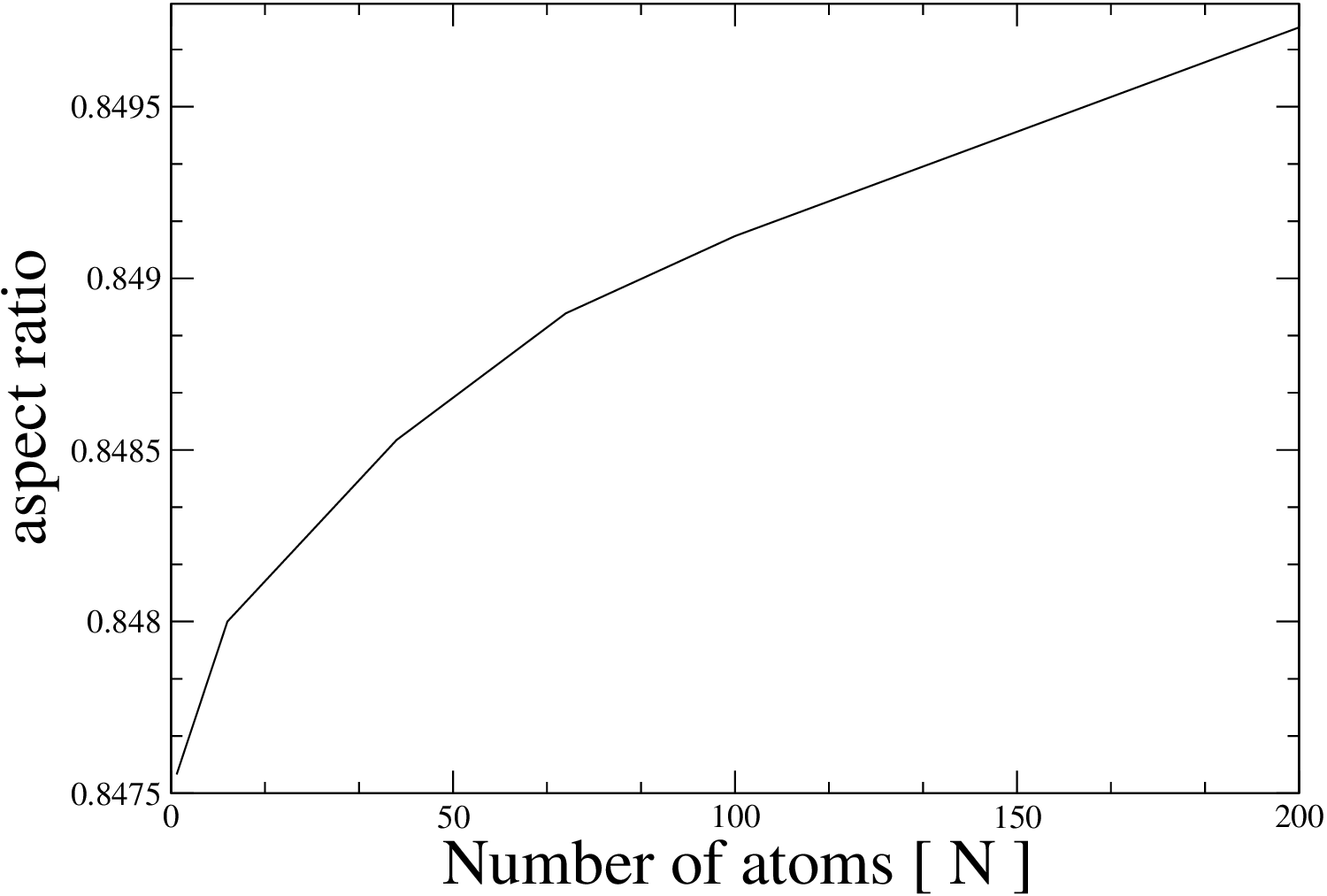}}
\caption{Aspect ratio in $li^7$ as a function of N.}
\vskip 1cm
 \end{figure}
\section{Conclusions:}
Numerical work with bare Feynman-Kac procedure employing modern computers was
reported[6] for the first time for few electron systems after forty years of
original work[5] and seemed to be real useful for calculating atomic ground
and excited states[7].  Tremendous success[9] in atomic physics motivated us 
to apply it to condensed matter physics. 
We have performed our calculations only with 200 particles.
Nonetheless we have been able to demonstrate some of the 'Holy Grails' of Bose 
Condensation, namely we have been able to calculate energy, chemical potential,
aspect ratio etc which reasonably agree with  existing results achieved through
other methods. The reason for this is that even with small number of particles
we have sovled many boson systems nonperturbatively and fully quantum 
mechanically. Within the numerical limitations our many body technique
is exact as solve the full Hamiltonian with realistic potentals. We have 
calculated spectrum of Rb and Li gas by considering  realistic
potentials like Morse potential, instead of conventional pseudopotentials
for the first time, thus allowing one to do the calculation exactly
as opposed to the case of $\delta $ function potential where it is calculated
perturbatively.

We have been successful in achieving a lower value for Rb in both isotropic and anisotropic cases and Li ground state than that obtained by Gross-Pitaevski 
technique[2]. As a matter of fact in all three cases our results stay
lower than those coming from GP.  For the isotropic case ( Case 1 with a system 
of 3-20 Rb atoms ) this does not contradict results of Lieb and Yngvason (LY)
[23] which states that energies obtained from  GP equation provide an lower 
bound to the actual ground state of a Bose gas with non-negative, finite range,
spherical two-body potential. This is due to the fact that mean field theory 
inherent in GP equation does not apply for system with small number of 
particles[11]. In the case of anisotropic case, LY theorem does not apply as 
theinteracting two body potential is non spherical in nature. Only general 
principle that applies here is  the energies coming from the GP technique used 
in ref[1] are variational estimates (results in this case are
obtained  by minimizing the energy functional corresponding to the Hamiltonian 
with mean field interaction instead of having some correlations explicitly.) 
and provide an upper bound for the diffusion Monte Carlo or similar exact 
calculations. This is again in conformity with Dyson's upper bound [24] for 
low density Bose gas.
We also found that Born approximation is valid at low energy and low 
temperature. Using this approximation we could produce results in all three
cases which favorably compare to those given in the litearture. However 
it is crucial to choose the paramters used in the potential correctly as 
these parameters control the strength of the interaction directly.
There are no hard and fast rules for how weak the intercation should
be as it was pointed out in the noble lecture[25]. '' It is  far more dilute
and weakly interacting than liquid helium super fluids, for example, but far
more strongly intercating than the nonintercating light in a  laser beam".

We have found an alternative to Gross-Pitaevski technique 
and other mean field calculations. The method is extremely easy to implement 
compared to mean field tecniques and our fortran code at this point consists of
about 270 lines. The simplicity in our many body theory is appealing as the 
mere ability to add, subtract and toss a coin enables one 
to solve the many body problem. We employ an 
algorithm which is essentially parallel in nature so that eventually we can 
parallelize our code and calculate thermodynamic properties of bigger systems
( of the order of 2000 atoms ) taking advantage of new computer architechtures. This work is in progress. We are continuing on this problem and hope that this 
technique will inspire others to do similar calculations. 

\newpage
{\bf Acknowledgements}:\\
Financial help from DST ( under Young Scientist Scheme 
(award no. SR/FTP/-76/2001 )) is gratefully acknowledged. The author would like
to thank Prof J.  K . Bhattacharjee, Indian Association for the Cultivation
of Science, India for suggesting the problem and many stimulating discussions 
and also Prof C. W. Clark of NIST, USA for suggesting very usuful references.
\end{document}